\documentclass[onecolumn,secnumarabic,amssymb, nobibnotes, aps, pre]{revtex4-1}

\usepackage{graphicx}
\usepackage{dcolumn}
\usepackage{bm}
\usepackage[usenames]{color}

\usepackage[all]{xy}
\usepackage{mathtools}
\usepackage{amsfonts}
\usepackage{amssymb}
\usepackage{amsmath}
\usepackage{natbib}
\usepackage{longtable}
\usepackage[toc]{appendix}
\usepackage[bottom]{footmisc}

\def \bea{\begin{eqnarray}}
\def \eea{\end{eqnarray}}
\def \be{\begin{equation}}
\def \ee{\end{equation}}

\def\({\left(} \def\){\right)}

\begin{document}

\title{An Effective Hydrodynamic Description of Marching Locusts}

\author{Dan Gorbonos$^{1}$\footnote{D.G and F.O. both contributed equally to this work.}, 
\setcounter{footnote}{0}
Felix Oberhauser$^{2,3\, a}$, Luke L. Costello$^{1}$, Yannick Günzel$^{2,3}$, Einat Couzin-Fuchs$^{2,3}$, Benjamin Koger$^{1}$ and Iain D. Couzin$^{1,2,3}$}

\affiliation{$^1$Department of Collective Behaviour, Max Planck Institute of Animal Behavior, 78464 Konstanz, Germany}
\affiliation{$^2$Centre for the Advanced Study of Collective Behaviour, University of Konstanz, 78464 Konstanz, Germany}
\affiliation{$^3$Department of Biology, University of Konstanz, 78464 Konstanz, Germany}

\begin{abstract}

A fundamental question in complex systems is how to relate interactions between individual components (``microscopic description'') to the global properties of the system (``macroscopic description''). Another fundamental question is whether such a macroscopic description exists at all and how well it describes the large-scale properties. Here, we address these questions using as a canonical example of a self-organizing complex system – the collective motion of desert locusts. One of the world’s most devastating insect plagues begins when flightless juvenile locusts form ``marching bands''. Moving through semiarid habitats in the search for food, these bands display remarkable coordinated motion. We investigated how well physical models can describe the flow of locusts within a band. For this, we filmed locusts within marching bands during an outbreak in Kenya and automatically tracked all individuals passing through the camera frame.  We first analysed the spatial topology of nearest neighbors and found individuals to be isotropically distributed. Despite this apparent randomness, a local order was observed in regions of high density with a clear second neighbor peak in the radial distribution function, akin to an ordered fluid. Furthermore, reconstructing individual locust trajectories revealed a highly-aligned movement, consistent with the one-dimensional version of the Toner-Tu equations, which are a generalization of the Navier-Stokes equations for fluids, used to describe the equivalent macroscopic fluid properties of active particles. Using this effective Toner-Tu equation, which relates the gradient of the pressure to the acceleration, we show that the effective ``pressure'' of locusts increases as a linear function of density in segments with highest polarization (for which the one-dimensional approximation is most appropriate). Our study thus demonstrates an effective hydrodynamic description of flow dynamics in plague locust swarms.
\end{abstract}

\maketitle
\section{Introduction}


The behavior of large animal collectives can be described through individual-level mutual interactions (this is termed a ``microscopic description''), much like describing a material through the interactions between its molecules. For fluids there is a complementary large-scale description in the continuum limit, the Navier-Stokes equations, which characterize fluids in terms of macroscopic parameters. An important question is whether large animal collectives, which appear to have an animate fluid-like character, can be effectively described by equivalent hydrodynamic descriptions of macroscopic properties such as pressure and viscosity~\cite{julicher2018hydrodynamic}. 


From the perspective of physics, large animal collectives can be thought of as a form of ``active matter'', a general name for agents or constituents that can convert stored energy into directed motion, forces, and shape deformations ~\cite{TONER2005170,RevModPhys.85.1143,shaebani2020computational}. It is not evident that an effective macroscopic description of active matter always exists and if such a description can correctly capture the systems main macroscopic properties. Living systems provide the preeminent example of active matter and spontaneous emergence of ordered movement in a group of many organisms. Here, we study marching locusts, that offer an intriguing opportunity to delve into the dynamics of such large animal collectives and a system where we can examine these questions about the macroscopic effective description.  We note that such collective behaviors appear, in addition to locust swarms, in many other collectives including migrating cell sheets~\cite{trepat2012cell}, and large aggregates of mobile group-living animals~\cite{ACKERMANN20211}. 

Locusts, a form of grasshopper, are considered a notorious agricultural pest due to their ability to form swarms that can cause widespread crop damage. Locust swarms can extend over several hundred square kilometers with up to 800,000 individuals per hectare, threatening the food supplies of millions~\cite{sivanpillai2016biological}. The formation of swarms begins when flightless juvenile locusts form so-called ``marching bands''. Facing challenges in obtaining sufficient nutrients in arid areas and under the risks of cannibalism ~\cite{bazazi2008collective, chang2023chemical}, marching bands make long-distance diurnal marching bouts. Field observations describe bands moving in a highly coherent direction and maintain relatively defined structures of columnar streams and planar fronts~\cite{kennedy1945observations, buhl2011group, buhl2012using}. However, due to the unpredictable nature of locust plagues and restricted access to infested areas, access to high-resolution field data is limited (with a few notable exceptions including detailed descriptive studies of~\cite{kennedy1945observations} and~\cite{ellis1957field, ashall1962studies} on desert locusts, and~\cite{clark1965sexual, clark1969field, clark1969night, buhl2011group} on Australian plague locusts). With recent advances in computer vision and automated animal tracking (\textit{e.g.}, ~\cite{tinevez2017trackmate, mathis2018deeplabcut, walter2021trex, koger2022multi}), large quantities of trajectory data can now be extracted automatically (\textit{e.g.,} see~\cite{buhl2011group} for manual, and~\cite{weinburd2021anisotropic} for automated tracking of comparable data sets) for fine-scale analyses of the structure of marching bands starting from the motion of individual animals~\cite{weinburd2021anisotropic}. 

Laboratory and modelling studies advance our understanding of locust swarming behaviour by deciphering local interaction rules that govern collective locust motion~\cite{ buhl2006disorder,  bazazi2008collective}, foraging decisions~\cite{gunzel2023information, dkhili2017self} and global group structures/dynamics~\cite{guttal2012cannibalism, buhl2012using, bernoff2020agent}. 
Supported by empirical evidence that non-cooperative interactions such as cannibalism can drive locust collective motion~\cite{ bazazi2008collective}, the principal models of attraction and repulsion~\cite{ buhl2006disorder} have been extended to anisotopic interactions of escape and pursuits, allowing the reproduction of a wide range of collective dynamics and social phenotypes~\cite{guttal2012cannibalism, romanczuk2009collective, romanczuk2012swarming}. 


Here we want to examine whether a macroscopic description of marching locusts exists and to what extent this description is successful.
Since flightless juvenile locust swarms move, predominantly, in two dimensions, and are so large, they provide a meaningful system with which to ask questions about the continuum limit of active matter (in two dimensions). Systems of active matter are far from thermodynamic equilibrium as a result of the constant energetic input (chemical energy from nutrients in the case of biological active matter or external fields in the case of engineered active matter), the energy is not conserved and therefore time reversal symmetry is broken. However, as shown below, some symmetries and conservation laws are still valid, like invariance under rotations and translations, and mass conservation~\cite{TONER2005170}.

One of the first models that became very popular to consider the physics of collectives, due to its simplicity and universality was the Vicesk model~\cite{PhysRevLett.75.1226}. This model is based only on local alignment interactions between the individuals and demonstrated how such interactions can account for the spontaneous emergence of ordered movement at large length-scales ~\cite{vicsek2012collective,ginelli2016physics}. The prediction of this model of a transition from disordered to ordered movement above a critical density was confirmed in locusts~\cite{buhl2006disorder}. Here we want to take another step forward. Since living organisms exhibit an overall complexity that should be, in principle, described by a large number of variables, it may seem counter-intuitive that such simple models like the Vicesk model, may be appropriate descriptors of macroscopic collective properties, such as long-range alignment. However, much as understanding the properties of fluids does not depend on an atomic-level description, it may be possible for active matter to also consider an effective theory to describe its movement using a continuous media approach that takes into account a small number of variables.  The continuous media approach gives a macroscopic long-wavelength description of the system by writing down the most general continuum equations of motion consistent with the symmetries and conservation laws of the system. Such a description of systems works for a large number of individuals while taking averages of the relevant quantities over long periods of time, assuming that the changes are slow in time and small in space (over the characteristic times and sizes of the system). Importantly, this approach does not depend on a specific microscopic description but rather introduces constraints on any microscopic model. Those constraints are introduced through the functional dependence of the variables in the continuous description; the velocity field $\vec{v}$ and the density $\rho$. In addition, there are phenomenological parameters in those equations, whose numerical value will depend on the detailed microscopic description of the interactions in the system. We can measure those parameters from the data and any new microscopic model in the future should provide us with those values.

The Navier-Stokes equation is the most general equation that describes fluids up to second order derivatives with respect to the spatial coordinates and is compatible with rotational and translational symmetry (for example see~\cite{LandauFluid}). 
Motivated by the inherently discrete Vicsek model, Toner and Tu generalized the Navier-Stokes equation to accommodate for the inherent properties of active fluids (absence of Galilean invariance; that is the laws of motion may not be the same in all inertial frames of reference) and to include an alignment force.  The marching locusts that are analyzed in this paper provide us with a unique opportunity to evaluate how the continuous media approximation works in general and the Toner-Tu equation~(\ref{TonerTu}) in particular since fits to the Toner-Tu equation are usually studied in simulations or artificial active matter and not in living matter, except for some unique works on starlings~\cite{ballerini2008empirical,mora2016local}. In this paper we will concentrate on the leading order terms in derivatives of Eq.~(\ref{TonerTu}) and we leave the sub-leading effects, e.g. viscosity, for future studies.

\begin{figure}[htb]
\centering
\includegraphics[width=0.8\linewidth]{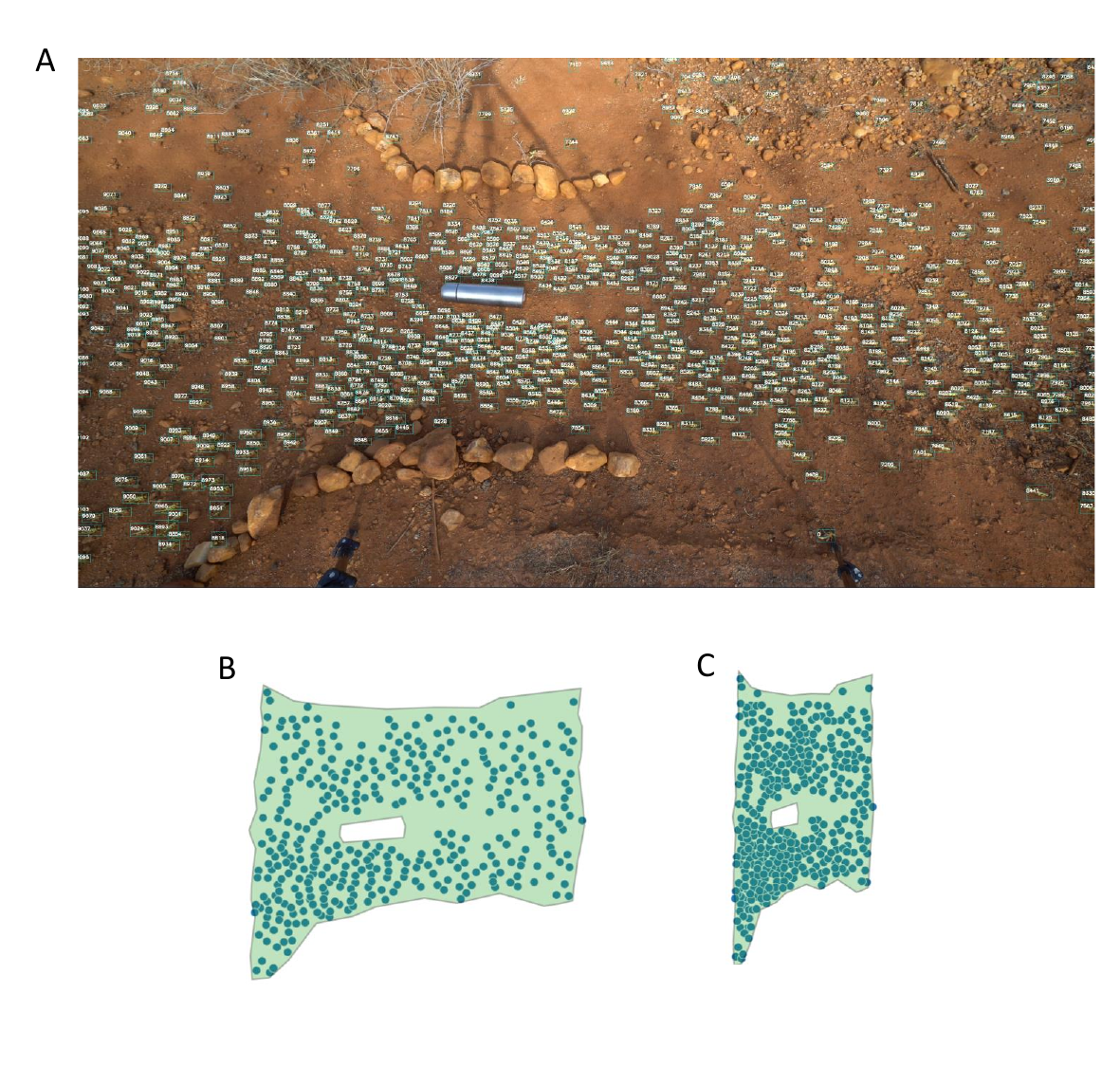}
\caption{(A) Example of a frame in a typical video frame. Each detected locust is surrounded by a bounding box (turquoise) and assigned a unique ID (white).  (B) Moving locust groups were grouped together in an $\alpha$-shape (green polygon, see appendix \ref{AppendixGlobalPacking}). Each point represents a locust. Due to the ellipsoid body shape of locusts, a correction was made by adjusting the distance along the moving direction by the body length/width ratio. (C) This reshaping led to the condensed shape shown, which was used to control for body shape induced phenomena. \label{fig:ExampleImage}
 }
\end{figure}

The data for the analysis presented here were collected in 2020 during a major locust outbreak in northern Kenya. A ``funnel'' was constructed using multiple rocks within a vast swarm, and a  places as an obstruction (see Fig.~\ref{fig:ExampleImage}A). The obstacle and the ``funnel'' together were designed in order to maximize the spatial variation of the density.
Recordings of the locusts moving through this configuration were made by filming an area of 3x2 meters approximately, directly down, and the individual trajectories of all locusts passing through were tracked. The first analysis that we present here is that of nearest neighbors distribution and radial distribution function (RDF). The RDF suggests that the structure is fluid-like and motivates us to consider the continuous media approximation, namely the Toner-Tu equations. We extracted the trajectories in two long strips on both sides of the obstacle and found long time periods (130-300sec) in which the movement is in a steady state. It means that the spatial average density and velocity (over a strip) are approximately constant up to small variations. This allows us to ignore terms with explicit time derivatives in the Toner-Tu equations and simplify the analysis. Moreover, it allows us to look at long time averages and compare quantities, like density and velocity, at different points along the strip. Only when we have steady state time average is meaningful for a specific point in space. 


In addition to the consistency of the continuous media approach, we found a simple equation of state that holds in the system - a simple relation of the hydrodynamic pressure with the density for a significant range of densities. Apriori we do not expect to find that an equation of state holds for active fluids. We usually find equations of state in thermal equilibria and the current system is far from it. In addition, the general functional form of the pressure is unknown, but based on the data, we can find its approximate dependence on the density. Some recent studies of an equation of state for active matter and systems far from equilibrium include simple models of active particles and active colloids~\cite{PhysRevX.5.011004, PhysRevLett.113.028103, C4SM00927D, PhysRevX.5.011004,PhysRevLett.121.098003,levis2017active, das2019local,
pirhadi2021dependency,dogra2022universal} and only a few examples for biological systems~\cite{silverberg2013collective,sinhuber2021equation, reynolds2021understanding}. The analysis below shows, a strong linear relationship between the pressure and density for a wide range of densities, namely
\be
P \propto \rho. \label{press}
\ee
The linear dependence is similar to ideal gas-like behavior. A similar resemblance to ideal gas has been observed in human crowd dynamics (see, for example,~\cite{silverberg2013collective}). Additionally, a non-linear generalized ideal gas law has been used to describe the thermodynamics of swarming midges~\cite{sinhuber2021equation, reynolds2021understanding}. It suggests that the thermal equilibrium may be a good approximation in those regimes. Then the equation of state gives a predictive power for a wide range of densities and boundary conditions that were not considered here. 

This equation of state (Eq.~\ref{press}) for the hydrodynamic pressure motivated us to look in addition at the mechanical pressure, which is related to deformations and not to the kinetics of the fluid. In thermal equilibrium they are balanced and equal. Therefore, we do not expect them to be equal when the system is far from thermal equilibrium (see also~\cite{pressure}), as in the current active system, which is studied here. Nevertheless, we find a similar linear relationship between the pressure and density for a similar range of densities.


\subsection{The Toner-Tu Equations}
The Toner-Tu equation~\cite{PhysRevLett.75.4326, PhysRevE.58.4828} can be written in the following schematic form:
\be
\partial_{t}\vec{v}+\lambda_1\,(\vec{v}\cdot\nabla)\vec{v}+\lambda_2\(\nabla\cdot\vec{v}\)\vec{v}+\lambda_3\nabla(|\vec{v}|^2)=\vec{v}F(|\vec{v}|)-\nabla P(\vec{x})+D_{B}\nabla\(\nabla \cdot \vec{v}\)+D_{T}\nabla^2\vec{v}+D_{2}(\vec{v}\cdot\nabla)^2\vec{v}+\sum_{i}\vec{\textit{F}}_i, \label{TonerTu}
\ee
where $\lambda_1,\lambda_2,\lambda_3$ are coefficients of advection/convection-like terms. This is the most general form of such terms. Under the requirement of Galilean invariance, which means that  the laws of motion are the same in all inertial frames of reference, the equation is reduced to the Navier-Stokes equation when $\lambda_1=1, \lambda_2=\lambda_3=0$. $F(|\vec{v}|)$ is an effective restoring force that keeps the speed constant at $v_0$, i.e., $F(v)$ is a one-dimensional function so that $F(v_0)=0$. This preferred speed breaks Galilean invariance for active matter. $P(\vec{x})$ is the hydrodynamic pressure whose gradient gives the total acceleration of the fluid particles (the local and the convective accelerations). This pressure is identified as the trace of the bulk stress tensor, whose microscopic definition is in terms of momentum fluxes~\cite{allen1987computer}. As mentioned above, for active matter this definition does not coincide in general with the mechanical pressure that is described below~\cite{pressure}. $D_{B},D_{T},D_{2}$ are coefficients of the viscosity-like terms. They contain the second order derivatives with respect to the spatial coordinates and are thus important when there are rapid spatial variations. In our case, they are subleading to the convective terms (which are first order in the derivatives) since we concentrate on slow spatial variations in the long wavelength approximation. We will see that in our approximation $\Delta v/\Delta x<1$ and therefore $\Delta v/\Delta x^2<\Delta v/\Delta x$, which justifies the hierarchy of the derivative expansion. The $\vec{\textit{F}}_i$s are all of the additional external forces that may act on the fluid.

\section{Local order and the distribution of neighbors}

In order to better understand the spatial structure of locust bands, we started by computing the radial distribution function of nearest neighbors. From the tracked data (see Fig.~\ref{fig:ExampleImage}A) we measured the nearest neighbor distribution from all locusts grouped by an $\alpha$-shape polygon (see appendix \ref{AppendixGlobalPacking}). 



\begin{figure}[htb]
\centering
\includegraphics[width=0.95\linewidth]{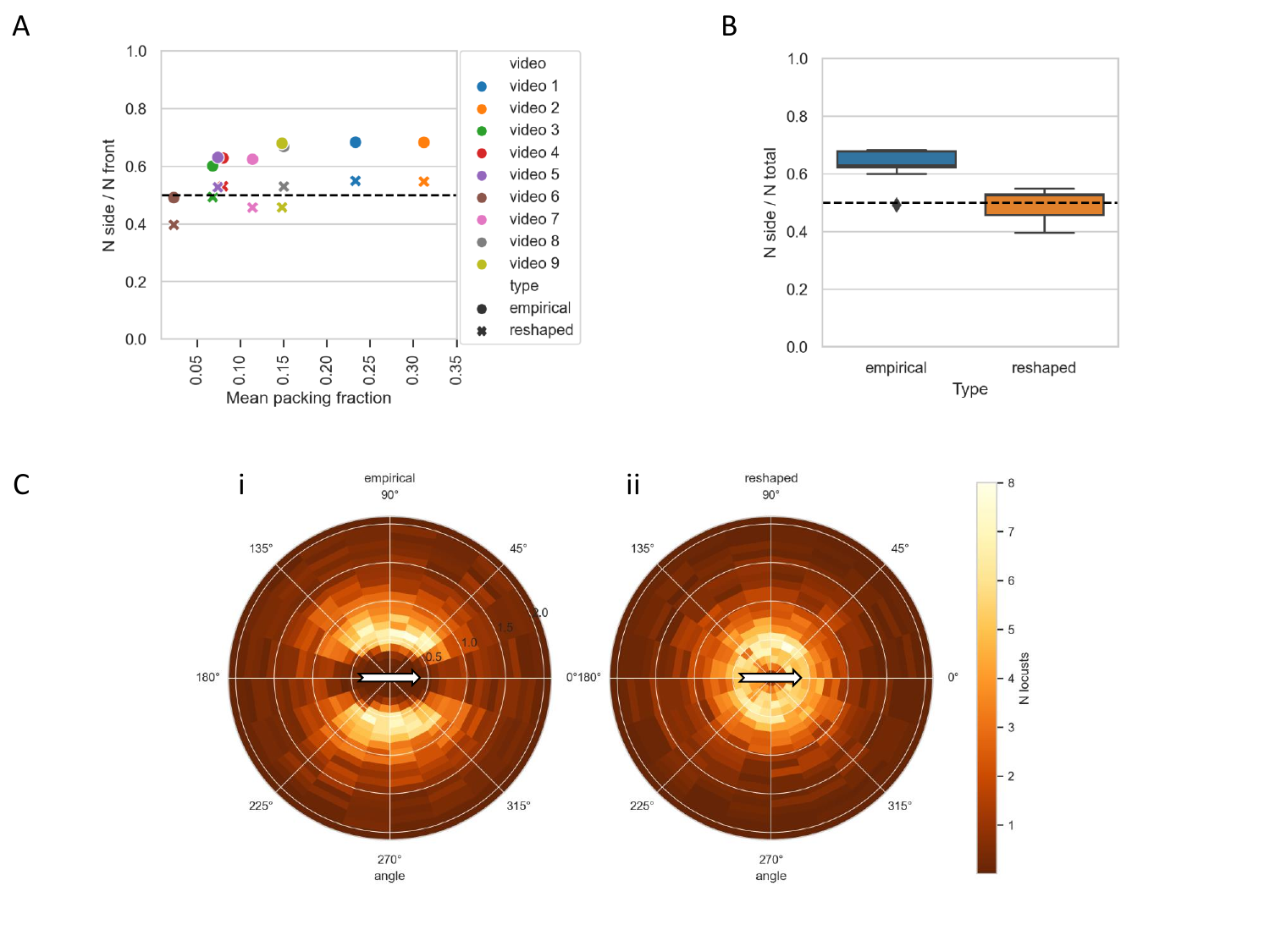}
\caption{\label{fig:NN_dist}
(A) Anisotropy (number of nearest neighbours on sides / total) of each recording for empirical data (full circles) and scaled data (crosses), which accounts for locust body shape. The overall pattern remains stable across recordings despite varying packing fractions, with a higher numbers of locusts on the sides in empirical data. In rescaled data, this anisotropy disappears and approaches isotropy (dotted line).
(B) Anisotropy for empirical and rescaled data over all recordings.  As in (A), the nearest neighbour distribution approaches isotropy once data is rescaled. The areas are defined in appendix \ref{AppendixQuadrants}.
(C) Heatmap of nearest neighbour distribution across all recordings within 2 body lengths around focal locust. Anisotropy is evident in empirical data (left), but disappears in rescaled data (right). Concentric white circles represent distance from focal locust in half body lengths. 0° is moving direction, also indicated by a white arrow.
 }
\end{figure}

As individuals defining the $\alpha$-shape's borders are surrounded by more empty space and thus have overall larger nearest neighbour distances, we corrected for this edge effect by applying a simplified Hanisch correction~\cite{Hanisch,cavagna2008starflag}: For each locust, both the distance to the shape border and the nearest neighbour distance were calculated. Only pairs where the distance to the nearest neighbour was smaller than the distance to the border were included, thus preventing an edge bias. These calculations were conducted once per second and used to calculate heatmaps of the nearest neighbour distributions (Fig.~\ref{fig:NN_dist}C. 

To quantify potential anisotropy, we binned the distributions in angular quadrants: front and back (135°-225°, 315°-45°), left and right (45°-135° , 225°-315, see Fig.~\ref{fig:NN_dist}A). We  calculated then the proportion of neighbors on the sides out of the total number of neighbors in all quadrants, $N_{sides}/N_{Total}$, and took this ratio as a measure of the isotropy (Fig. \ref{fig:NN_dist}A and Fig.~\ref{quadrants} in appendix \ref{AppendixQuadrants} for the definition of the quadrants). For an isotropic distribution this value is 0.5.

In the raw data (Fig.~\ref{fig:NN_dist}Ci) there is a clear anisotropy, with near neighbors more likely to be found on the sides than the front or back. This, however, might be an artifact of the animal's body shape; locusts are not circular but have an elliptical body shape. Elongated shape allows nematic ordering which could be evolutionary adaptive for collective movement. Because of the elongated shape, it is more likely to measure a near neighbor on the side. We can correct for this bias by rescaling with the aim being to take the elliptical body shapes and transform them to circles. In this data set the alignment is high since the locusts are all marching the same direction through a channel, and the velocity vectors (and hence the orientations) of all individuals lie along roughly the same direction. Because of this alignment we can apply the same transformation to all individuals with approximately the same result. Since we rotate our data so that the mean velocities are always along a principle direction we can preform this rescaling operation by simply multiplying the coordinates along one principle direction by a scale factor. The scale factor is derived from the aspect ratio of the bounding boxes from the tracking. We then used this scale factor to correct the shape by squeezing the long x-axis to obtain a 1:1 ratio. This data is referred to as ‘rescaled’ and an example of the result of this process is shown in Fig.~\ref{fig:ExampleImage}B-C. The nearest neighbor distribution over the rescaled data is roughly uniform (Fig.~\ref{fig:NN_dist}Cii), indicating that there is no preference for side to side distribution (Fig.~\ref{fig:NN_dist}B). The effect of rescaling is consistent across all recordings (Fig.~\ref{fig:NN_dist}A). %

\begin{section}{The radial distribution function} 
Since the distribution of neighbors (in the corrected coordinates) is radially symmetric we can use the radial distribution of neighbors as a way of quantifying the physical structure of the group. The radial distribution function (RDF) is a common way of quantifying the local structure of materials and small molecules. The RDF measures the local density of neighbors relative to a focal particle (or in our case individual), averaged over all particles, as a function of radius. On a perfectly ordered lattice the RDF will display sharp peaks at radial distances corresponding to the locations of neighbors. The shape of the RDF can provides qualitative information about the state of a material system. For example, as a system heats up, the particles oscillate about their equilibrium positions and the sharp peaks in the RDF which would be observed for static particles become more diffuse~\cite{ding2014temperature}. More diffuse peaks indicate higher disorder. 

\begin{figure}[htb]    \centering
    \includegraphics[width=0.7\linewidth]{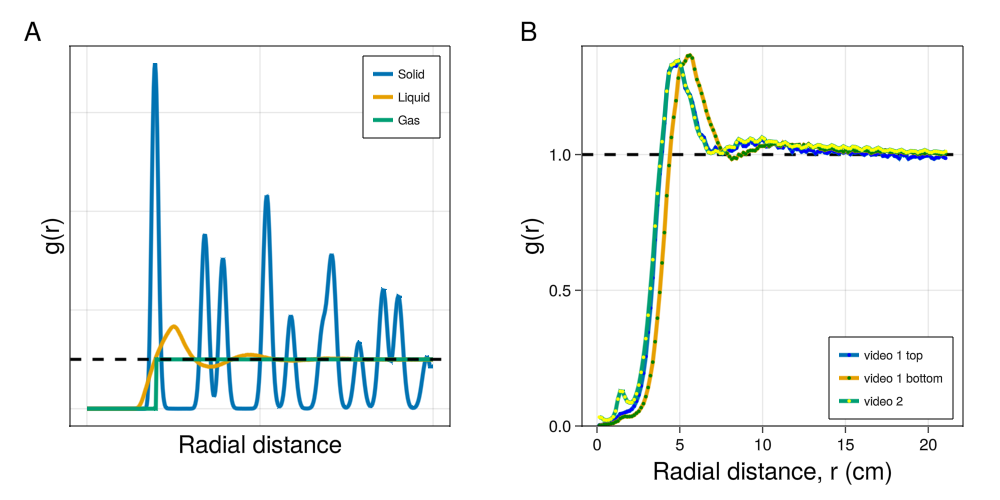}
    \caption{Panel A is a cartoon of the radial distribution function for Lennard-Jones disks in the solid, liquid, and gas phases. The radial distribution functions for the highest density regions in videos 1 and 2 are shown in B. The dashed line is $y=1$, which the RDF should converge to at long distances.}
    \label{fig:RDF}
\end{figure}

For our data we calculate the RDF as a way of qualitatively comparing the distribution of marching locusts to the expected RDF for various states of matter to understand what physical state the marching locusts most resemble. The RDF is calculated as

\begin{equation}
    g(r) = \left< \frac{n(r)}{a(r)} \right>
\end{equation}

\noindent where  $n(r)$ is the number of individuals in an annulus of radius $r$ relative to a focal individual, $a(r)$ is the area of this annulus (thus $n(r)/a(r)$ is the density of individuals in the annulus of radius $r$) and the brackets denote the average over all individuals. This is computed for specific regions in two of the recordings where we expect densities to be the highest. The focus is on regions which are above and below the obstacle. In these regions the densities are high but the channels are narrow so the number of individuals and their neighbors is somewhat low. Further, many individuals are close to edges, which strongly skews the RDF. To mitigate this a supercell is constructed from each of the regions. This is effectively considering the region boundaries to be periodic. With this consideration we eliminate the effect of the edges without losing the contributions of the individuals that lie near an edge. There calculation are all performed on the rescaled data.

We focus the analysis on the two videos with the highest density, video 1 and video 2, (significantly higher packing fraction than the other recordings, see Fig.~\ref{fig:NN_dist}A and Fig.~\ref{Packing Fraction} for the comparison of the packing fraction of all the nine recordings) which are sufficient for time averaging and the continuous media approximation (that we consider below) to hold. 

The placement of the elongated obstacle varies across the two recordings. In video 1 the channel is nearly symmetric with the elongated obstacle placed in the center. In the other the elongated obstacle is placed asymmetrically, closer to one side of the channel. Because of this placement there is a wide channel and a narrow channel. We only include the wide channels in calculating the RDF for this recording. For the nearly symmetric recording we calculate the RDF in both channels (labeld ``top'' and ``bottom'' in Fig.~\ref{fig:RDF} B).

The calculated RDFs are shown in Fig.~\ref{fig:RDF}. There is a clear first neighbor peak sometimes followed by a weak second neighbor peak. This second peak only appears in the highest densities and indicates some qualitative similarity to a liquid crystal~\cite{shrivastav2021self} especially as compared to random packed hard spheres~\cite{yuste1991radial} for instance. The presence of some amount of order in the RDF lends support to our consideration of this system as fluid like.
\end{section}

\section{The Hydrodynamic Pressure}

The elongated obstacle (the bottle) splits the stream into two narrower streams. Together with the funnel shape of the boundaries it contributed to the spatial variation of the density and speed. Our goal is to extract values of velocity and density on a range as wide as possible at various points in space in order to check the validity of the hydrodynamic effective description and an equation of state like the one in Eq.~\ref{press}. For this purpose we defined two long strips on both sides of the obstacle, extracted trajectories, and looked at long periods of time in which the average density and velocity are approximately constant. Only then spatial differences are meaningful. The strips are shown in Fig.\ref{Second}A and defined in detail in appendix \ref{AppendixSteady}.
In the following analysis we compare the dynamics of the locust movement in the continuum limit with the Toner-Tu equation. To complete the full continuum description of the Toner-Tu equation, we require continuity equation which is a mass conservation statement:
\be
\partial_t \rho-\nabla(\rho \vec{v})=0. \label{continuity}
\ee

From equations~(\ref{TonerTu}) and~(\ref{continuity}) we see that we have three requirements for a steady flow; the density, the speed and the direction of motion should all remain constant. We consider the system in a time segment to be in a steady flow if for a quantity $M$ its standard deviation is less than 10 percent of its mean value. We show in appendix \ref{AppendixSteady} how they are satisfied for long periods of time in the two strips of video 1 (330 sec approximately for the bottom strip and 120 sec for the top strip).

\subsection{The one-dimensional approximation}

\begin{figure}[!htb]
\centering
\includegraphics[width=0.8\linewidth]{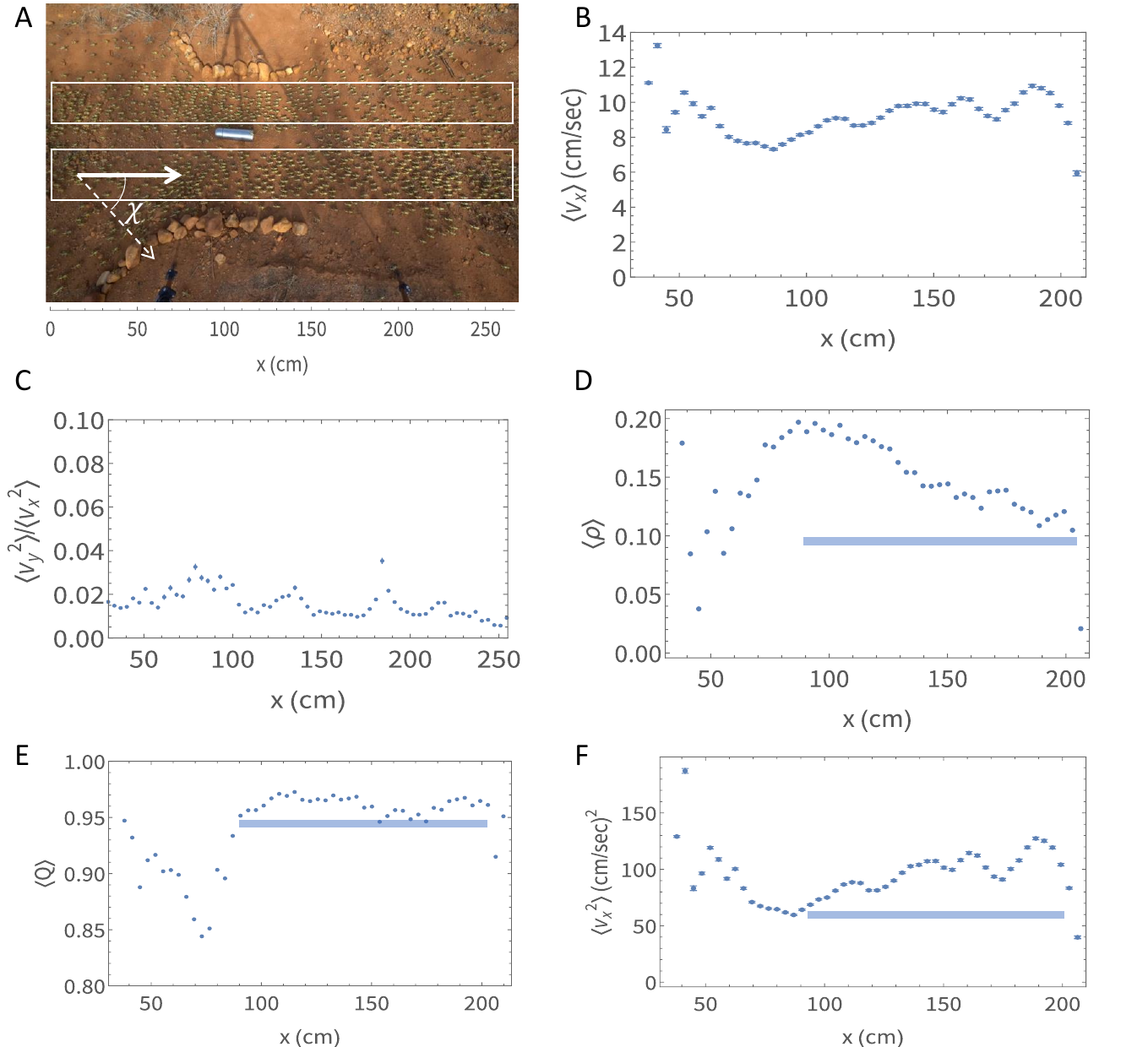}
\caption{\label{Second}
(A) A snapshot of video 1 with the areas took for analysis marked by white boxes. All panels show analysis of quantities averaged over time for the bottom strip for which the definition of the alignment angle $\chi$ and horizontal distance ruler are marked (in cm). (B) The horizontal component of the velocity $v_x$ as a function of the horizontal coordinate $x$ (average over time).(C) the ratio of the velocity components squared as a function of $x$. (D) The density $\langle \rho \rangle$ (packing fraction) as a function of the horizontal coordinate $x$. (E) The alignment order parameter as a function of $x$. The horizontal blue line denotes the segment where the alignment is maximal $ \langle Q \rangle >0.95$. (F) The horizontal component of the velocity $v_x$ squared as a function of $x$.  }
\end{figure}

In this section we only consider time segments when our steady flow conditions are satisfied. Then we can look at the average velocity and density over time, compare different points in space, and concentrate on the effect of the geometry of the environment on the motion. The experimental arrangement in the field created an effective two-dimensional funnel that focuses the marching locusts towards the elongated obstacle (the bottle in the recording) around which the density is higher and the speed is reduced. Past the obstacle the locusts spread out, lowering their density but increasing their speed. Thus the movement can be divided into three sections relative to the obstacle in the middle (see Fig.~\ref{Second}A-D and appendix \ref{AppendixGlobalPacking} for packing fraction in the three sections). Note that every point in the plots of Fig.~\ref{Second} is obtained by averaging over a long time period in which the strip is in steady state.

\begin{figure}[!htb]
\centering
\includegraphics[width=0.4\linewidth]{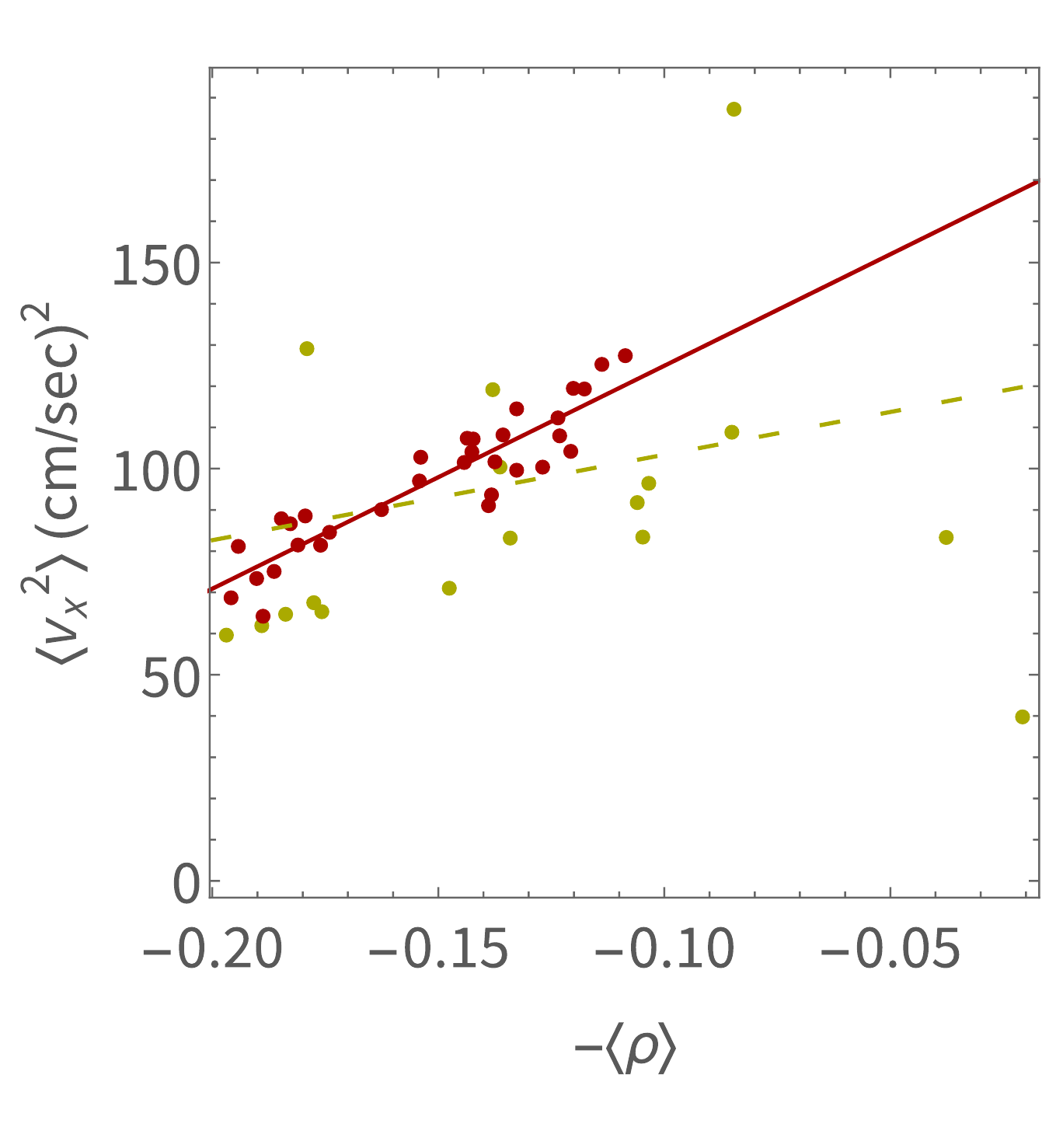}
\caption{\label{Second2} $v_{x}^2$ as a function of -$ \langle \rho \rangle$ for the bottom strip. The red points correspond to strong alignment $ \langle Q \rangle >0.95$ (from Fig.~\ref{Second}) and the yellow points are the rest. The linear fit of the points with strong alignment is in red and $R^2=0.84$. When we include all the points in the fit we get the dashed yellow line with $R^2=0.11$.}
\end{figure}

The funneling effect of this geometrical structure makes it attractive to consider a one-dimensional approximation as the first approximation for the dynamics of the system. We then consider mean values of the variables along narrow vertical strips at constant $x$ and, this way, neglect their variation with respect to the vertical coordinate $y$. Within this approximation, the alignment force becomes a lower order effect relative to the convective acceleration, and the Toner-Tu equation~(\ref{TonerTu}) is reduced to the following:
\be
\partial_t v_x+\(\lambda_1+\lambda_2\)v_x\partial_x v_x+\lambda_3\partial_x\(v_x^2+v_y^2\)=-\partial_x P.
\ee
We look at strips whose width is 50 pixels (roughly 4 cm, the average length of a locust), since this is the relevant scale for spatial variations. In Figures \ref{Second}B and  \ref{Second}D we see the density and the horizontal component of the velocity $v_x$ as a function of the horizontal distance, $x$, where the angle brackets represent averages over time.

When we look at the ratio $v_y^2/v_x^2$ as a function of the position on the horizontal axis, $x$, we find that it is less than 0.1 (Fig. \ref{Second}C) and thus the term $v_y^2$ can be neglected, and we get the following equation:
\be
\partial_t v_x+\lambda_{eff}\partial_x(v_x^2)=-\partial_x\,P, \label{reducedOne}
\ee
where
\be
\lambda_{eff}\equiv\frac{\lambda_1+\lambda_2}{2}+\lambda_3. \label{OneDimMin}
\ee
In addition, we find that in steady flow the local acceleration is negligible compared to the convective one. That is an aditional simplification of Eq.~(\ref{OneDimMin}). We can see that by taking the average over time of Eq.~(\ref{reducedOne}). From the integration of the first term of the left-hand side over a time segment $T$ we get the mean acceleration which is bounded by a small number for a long time of measurement ($\sim 300$ sec, see Fig.~\ref{Steady})
\be
|\frac{1}{T}\int_{0}^{T}\partial_t v_{x} dt|=\frac{|v_x(T,x)-v_x(0,x)|}{T}<\frac{max|v_x(T,x)|-min|v_x(0,x)|}{T}<\frac{10}{300}\ll 1
\ee
compared to the second term where $\lambda_{eff} \sim O(1)$ and $v_x^2\sim 60 (cm/s)^2$. In conclusion, we see from the data that in the one-dimensional approximation the convective acceleration is the dominant term in its contribution to the total acceleration. 

\subsection{The alignment order parameter}

If the assumption about the linear dependence of the pressure on the density in Eq.~(\ref{press}) is correct and the one-dimensional approximation is good enough, we get from Eq.~(\ref{reducedOne}) that the density should be linear with the velocity squared:
\be
-\rho \propto v_{x}^2. \label{densityVel}
\ee
However, the accuracy of this relation depends mainly on the accuracy of the one-dimensional approximation of the hydrodynamic equations. When all individuals are aligned, this linear relation should be valid. To check this we can quantify how much the system is aligned by introducing an alignment order parameter~\cite{RevModPhys.85.1143} which is the average of $\cos(2\,\chi)$ over all individuals
\be
Q=\overline{\cos(2\,\chi)},
\ee
where $\chi$ is the angle between the local velocity and the average direction of movement (see Fig.~\ref{Second}A). The overline denotes the average over all individuals. $Q=0$ is when all the particles are in random orientation and $Q=1$ when they are perfectly aligned. From the Toner-Tu equation~(\ref{TonerTu}), we expect that when the alignment is stronger, the one-dimensional approximation should be better and then the linear relation~(\ref{densityVel}) should hold. The average alignment over steady flow is given in Fig.~\ref{Second}E. In the bottom strip the alignment is maximal ($\langle Q \rangle ~>0.95$) for $90<x<200~cm$ (marked by blue lines in Fig.~\ref{Second}E, and for $100<x<175~cm$ in the top strip, see appendix \ref{AppendixTopStrip}). For those segments we then check the linear correlation of $-\langle \rho \rangle$ with $ \langle v_{x}^2 \rangle$ and show that it is stronger in the segments where the alignment is stronger. In Fig.~\ref{Second}D and F the segments where the alignment is strongest are marked according to Fig.~\ref{Second}E. In Fig.~\ref{Second2} we compare the linear fit ($ \langle v_{x}^2 \rangle$ vs. $- \langle \rho \rangle$) of the segment with strongest alignment (red line and red points) with the linear fit using all points (yellow dashed line). We see that indeed the linear fit is significantly better when only the region of strong alignment is included. This way we could sample the locust dynamics in the regions where the one-dimensional approximation is good enough and find that the relation in Eq.~(\ref{press}) indeed holds. The geometrical constraints (the funnel and the bottle) were good enough in creating such conditions, so that the one-dimensional approximation is successful. We obviously expect that the equation of state holds everywhere where the pressure is high enough and does not depend on the particular boundary conditions.

\begin{figure}[!hbt]
    \centering
    \includegraphics[width=0.5\linewidth]{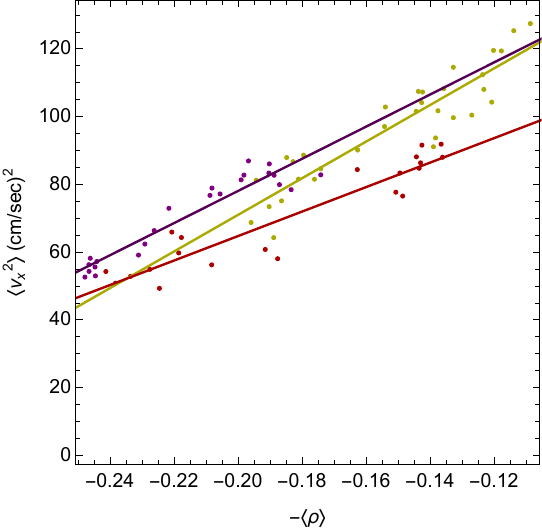}
   \caption{ \label{linearFig} $v_{x}^2$ as a 
   function of -$ \langle \rho \rangle$ for segments of strong alignment ($ \langle Q \rangle >0.95$) with linear fits for cases where they exist: The red points correspond to the top strip of video 1 ($R^2=0.88$), the yellow points for the bottom strip ($R^2=0.84$) and the purple points are for the bottom strip of video 2 ($R^2=0.9$). }
\end{figure}

We checked the same linear relation for more recordings but most of them, as evident in Fig.~\ref{Packing Fraction}, are at densities which are too low to contain a sufficient number of animals for analysis, and the polarization is lower as well (up to 0.8 on average). 
The results for the top strip of video 1 are given in  appendix \ref{AppendixTopStrip} (Figs.~\ref{SecondB1},~\ref{SecondB2}). In the second recording the top strip turned out to be too narrow for a significant stream. Therefore in Fig.~\ref{linearFig} we summarized values of the velocity squared vs. the density for three segments of strips with highest alignment ($Q>0.95$) and obtained very good linear fits.

\section{Mechanical Pressure}

\begin{figure}[!hbt]
    \centering
    \includegraphics[width=0.6\linewidth]{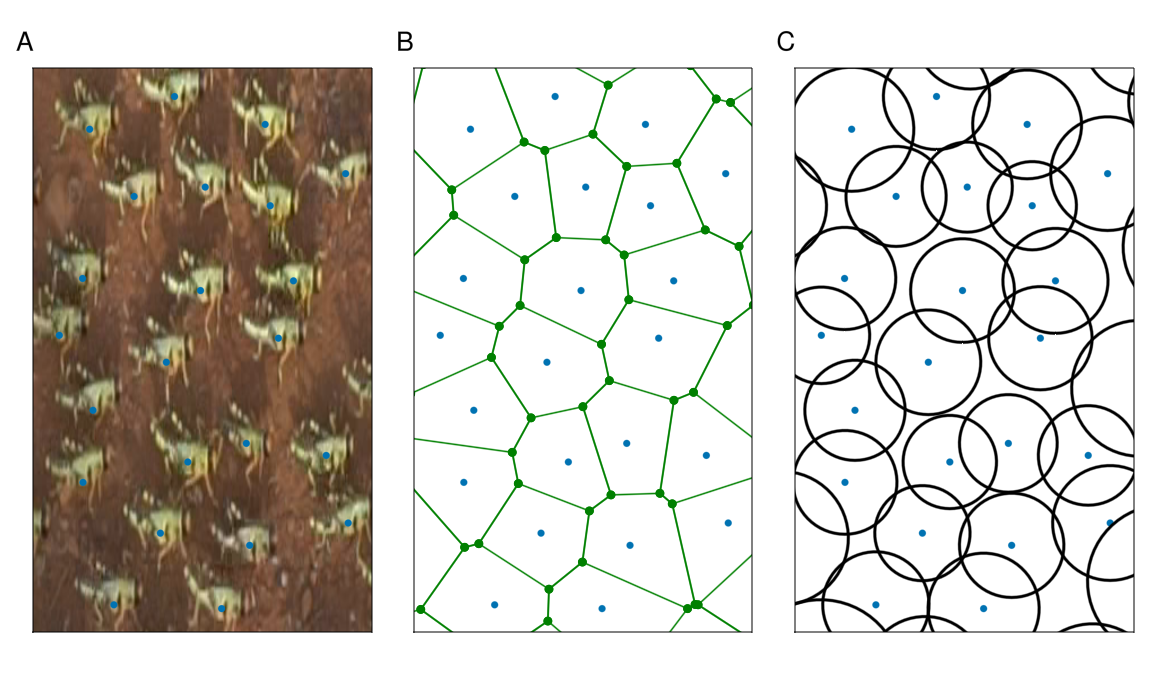}
    \caption{An illustration of the process for determining the radii, $r$, used in calculating the mechanical pressure. A snapshot of the locusts with the tracked centroids overlaid is shown in A. The image is stretched to match with the centroids in the rescaled coordinates. B shows the Voronoi tesselation over these points and C shows the circles generated from this tesselation. Each circle in C has an area equal to the corresponding Voronoi cell area.}
    \label{fig:Voro}
\end{figure}

In the previous section we develop a hydrodynamic definition of the pressure, which depends on the squared velocity. Additionally, we can define a static pressure through an analogy with the mechanical stress. The mechanical stress is defined as the force per unit area acting on the faces of a cubic volume element and is related to the strain (or deformation) by a generalized version of Hooke's Law~\cite{fung1966foundation}. This concept relates the force acting on an object (pressure) to the resultant deformation of that object. In the context of marching locusts we can think of the object as the area which an individual locust occupies and can calculate the force that must be exerted to shrink or grow that area relative to a fixed reference. We approximate the area occupied by a locust as the Voronoi area around the individual. Because the distribution of neighbors in the corrected coordinates is radially symmetric we can then approximate the deformation as acting on a circle of the same area as the Voronoi cell so that the mechanical pressure is radially isotropic (only acts on the radius of the circle) and one-dimensional. We can relate this to the hydrostatic stress, which is simply the isotropic component of the mechanical stress. We write this relationship as:

\begin{equation}
    \sigma_{ij} = E \epsilon_{ij}
\end{equation}

\noindent where $\sigma_{ij}$ is the mechanical stress, $E$ is the elastic modulus, and $\epsilon_{ij}$ is the strain. The isotropic component of the pressure acting on the radius of a circle is then $\sigma_{ii} = \sigma_{r} \equiv P_{mech}$, where $P_{mech}$ is our mechanical pressure and we get

\begin{equation}
    P_{mech} \propto \epsilon_{r} = \Delta r / r_m \label{eqn:P}
\end{equation}

\noindent where $r$ is the radius of the circle representing the area occupied by a single locust and $\Delta r = r - r_{m}$ is the change in the radius relative to a reference radius, $r_{m}$. To calculate the area occupied by a single locust we start by computing the Voronoi tessellation of a single frame, including the positions of all locusts. The area of each Voronoi cell is calculated and taken to be the total area around each locust. The radius, $r$, is then calculated for each individual as the radius of a circle which has the same area as the Voronoi cell. This process is illustrated in Fig.~\ref{fig:Voro}. In what follows we calculate the reference radius, $r_m$, per frame as the mean radius taken over all locusts in that frame.

\begin{figure}[!hbt]
    \centering
    \includegraphics[width=0.4\linewidth]{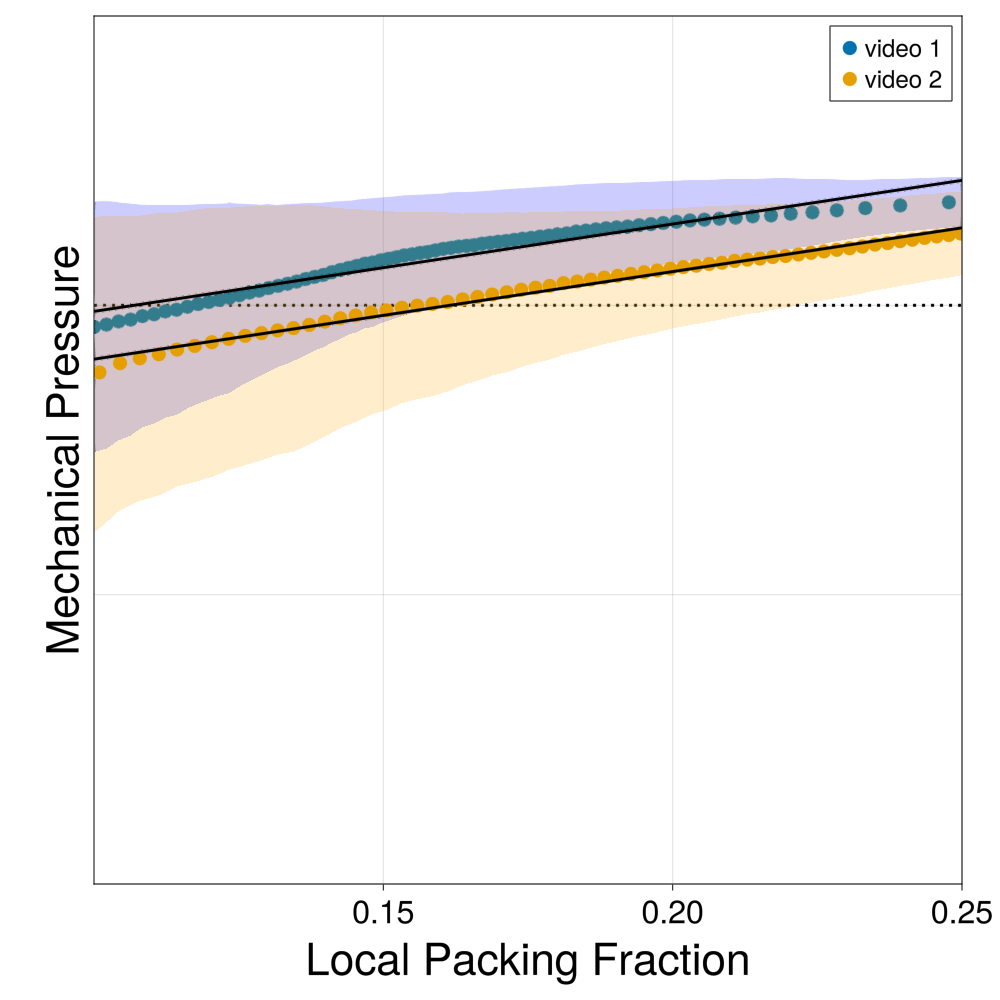}
    \caption{The mechanical pressure (Eq.~\ref{eqn:P}) vs local packing fraction (Eq.~\ref{eqn:rho}) for two recordings which span a high range of densities (video 1 and video 2). The points are the mean values of all points in fixed size bins (all containing the same number of samples) and the spread of the bands is the first standard deviation. The black lines are linear regressions on the displayed regions of each recording.  The $R^2$ values for videos 1 and 2 are 0.933 and 0.985 respectively. The dotted line is the line $y=0$. Note that the mechanical pressure here is defined dimensionless $P_{mech} = \Delta r / r_m$ (without an elastic constant) and is relative to the mean radius in a given frame.}
    \label{fig:P}
\end{figure}

In addition to the pressure we calculate the local packing fraction around each locust.

\begin{equation}
    \rho^{local}_{i} = \frac{\sum A_{j}}{A_0} \label{eqn:rho}
\end{equation}

\noindent where the summation is over all neighbors within a fixed radius, $r_0$ of a focal individual, $A_{j}$ is the area of the bounding box of a single neighboring locust and $A_0$ is the area of the circle with radius $r_0$. We choose $r_0$ to be roughly 9 locust body lengths and from this calculate the number of locusts in the that circle. This is because the maximum value of the local density distribution is constant for $r_0$ values above 6 (350px) body lengths but drops off, due to edge effects, for values higher than 12 (750px). Thus we choose a value between these bounds and from this calculate the number of locusts in that circle. The mechanical pressure vs packing fraction is shown in Fig.~\ref{fig:P}. There is, as with the hydrodynamic pressure, a strong linear relationship between the mechanical pressure and the local density. In general there is not always a well defined pressure in active systems~\cite{junot2017active, pressure}.  Here we find a consistent linear relationship between two definitions of pressure with the density, indicating that despite the active nature of this system the pressure is a well defined state function for a range of densities.

\section{Discussion}
In this paper we used tracked coordinate data obtained from recordings of marching locusts in the field. We find that, when accounting for the elliptical shape of the locust's body, neighbors are randomly distributed around a focal individual. Marching locusts, at least at the densities studied here, have no apparent preference for their position relative to their neighbors. Looking further at the distribution of neighbor distances we calculated the radial distribution function (RDF) for high density strips, finding that the RDF resembles that of fluid.

We then show that marching locust bands can be effectively described by a hydrodynamic description which is compatible with the Toner-Tu equations (the equivalent of the Navier-Stokes equations for active matter). In particular we find that an effective equation of state holds for a wide range of densities (Fig.~\ref{linearFig}). The fact that the linear relation in~(\ref{densityVel}) fits well to the data supports the relevance of the Toner-Tu equation~\ref{TonerTu} to describe the dynamics and supports the linear relation between the hydrodynamic pressure and the density (Eq. (\ref{press})). In addition we find another linear relationship between the mechanical pressure and local packing fraction over the same range of densities. We do not expect these relations to hold for all values of density but expect it to break down at low densities due to a change in the nature of mutual interactions of the locusts. At high densities the interactions are short range and predominant by effective hydrodynamic properties while at low densities long range interaction dominate.  In general the existence of an equation of state for the pressure is the exception for active matter~\cite{pressure}, and we find here that it exists for packing fractions in the range of 0.1-0.25. However, we see that although the linear relation in the hydrodynamic pressure holds for three separate strips (Fig.~\ref{linearFig}), the slope is different, even for two strips from the same recording. This slope is a phenomenological parameter that depends on the detailed microscopic description of the interactions in the system. It characterizes the specific material that is examined. Therefore in non-biological matter we expect it to be constant in contrary with the findings for this biological active matter. 
It therefore shows the ability of organisms to change effective macroscopic properties according to environmental conditions while preserving the same basic interactions which are reflected by this equation of state.

The effective macroscopic description that we found provides constraints on any microscopic model. Any assumptions on the functional form of the interactions between individual locusts should reproduce the equation of state that was found 
and explain how adaptation to different environmental conditions can be reflected by different phenomenological parameters.

\section{Material and Methods}

\subsection{Data collection}
Here, we analyze motion trajectories of juvenile marching bands (nymphs of 3rd to 5th stage) of desert locust \textit{Schistocerca gregaria}. Data were collected during a major locust outbreak in February and March 2020 in Samburu County, northern Kenya. Nine recordings of marching bands were used in the analysis with a total duration of 42.41 minutes and 160,164 assigned locust ids (note that the actual number of locusts might be lower as new ids were assigned once a locust disappeared for at least one second). All were recorded in 25~fps, with Sony Alpha II cameras placed on tripods facing the ground vertically, covering approximately an area of 3 x 2 meters. The cameras were set up in front of the marching bands to capture the locust flow once they reached the camera range without disturbance. The areas were all configured with a ``funnel'' built from rocks which forced the locusts into a narrow channel, where an elongated obstacle (a bottle) was placed. An example of a typical setup is shown in Fig.~\ref{fig:ExampleImage}. The nine videos were selected as to capture variation in the locust density and natural surroundings (such as fine sand or grainy sand substrate). While other recorded videos were used for obtaining training data, they were excluded due to reasons such as spontaneous change of direction, bidirectional flow, diagonal movement through the frame, propagating jumps or external perturbations.

\subsection{Visual tracking of animals in the field}
A training data set of 95 images was obtained by sampling 5 images spaced over the whole recording duration each for 18 recordings (one recording contributed 10 images). All locusts in these images were then manually annotated using the annotation software CVAT~\cite{cvat_ai_corporation_2023_7863887}. In total 48131 locusts were annotated. These annotations were used to train a pre-trained faster R-CNN model with a ResNet-50 + Feature Pyramid Network backbone to detect locusts in each image using Facebook AI’s detectron2 API (\texttt{faster\_rcnn\_R\_50\_FPN\_3x} ~\cite{wu2019detectron2}). The trained model was then used for inference on all images of the recording, yielding a bounding box (x, y, height, width) for each detected locust. Only detections with a certainty of at least 0.6 were kept and their quality displayed and reviewed in an annotated video file (see the video in the SI). Next, all the detections were linked using trackpy~\cite{allan_daniel_b_2021_4682814} with a search range of 100 pixels and memory of 25 frames (\textit{i.e.}, 1 second) using a k-d tree linking method. While creating these trajectories no interpolation took place; only an assignment of an estimated ID for each point was used. All IDs which were only present in one frame were subsequently removed. 

\subsection{Pre-processing of trajectory data}
\subsubsection{Filtering}
The tracked data was then further processed in preparation for data analysis. All detections within one locust's body length of the recording edges were removed to prevent tracking artifacts from appearing and disappearing locusts. Furthermore, stones, bushes, and other obstacles on the edges and the central obstacle were annotated so that all detections outside the experiment's boundaries could be excluded (see Fig.~ \ref{fig:ExampleImage}A). From the remaining trajectories, we estimated the median heading direction of all locusts and rotated the data so that the median heading direction was approximately aligned with the horizontal axis (locusts moving from left to right in the recording). 

\subsubsection{Assigning strips}
On the y-axis (orthogonal to the movement direction), two rectangle strips were assigned (see Fig.~\ref{Packing Fraction}A). The first spanning the shortest vertical path between the top boundary and the top of the obstacle annotation, the second between the shortest vertical path from the bottom of the obstacle to the lower boundary. Both strips have a margin of one body length distance from the boundaries. On the x-axis (along the movement direction), we separated the area into three regions according to the density of locusts over the whole duration of the recording. The first segment ends at the point of the highest observed density, the second segment ends at the first drop in density, and the third segment continues until reaching the right edge. These regions are shown in Fig.~ \ref{Packing Fraction}A.

\section{Acknowledgments}
D.G. and I.D.C. acknowledge support from the Office of Naval Research Grant N0001419-1-2556, Germany’s Excellence Strategy-EXC 2117-422037984 (to I.D.C.) and
the Max Planck Society, as well as the European Union’s Horizon 2020 research and innovation programme under the Marie Skłodowska-Curie Grant agreement (to I.D.C.; \#860949).

\section{References}
\bibliography{locust}

\appendix

\section{Steady Flow}
\label{AppendixSteady}

\begin{figure}[!hbt]
\centering
\includegraphics[width=0.9\linewidth]{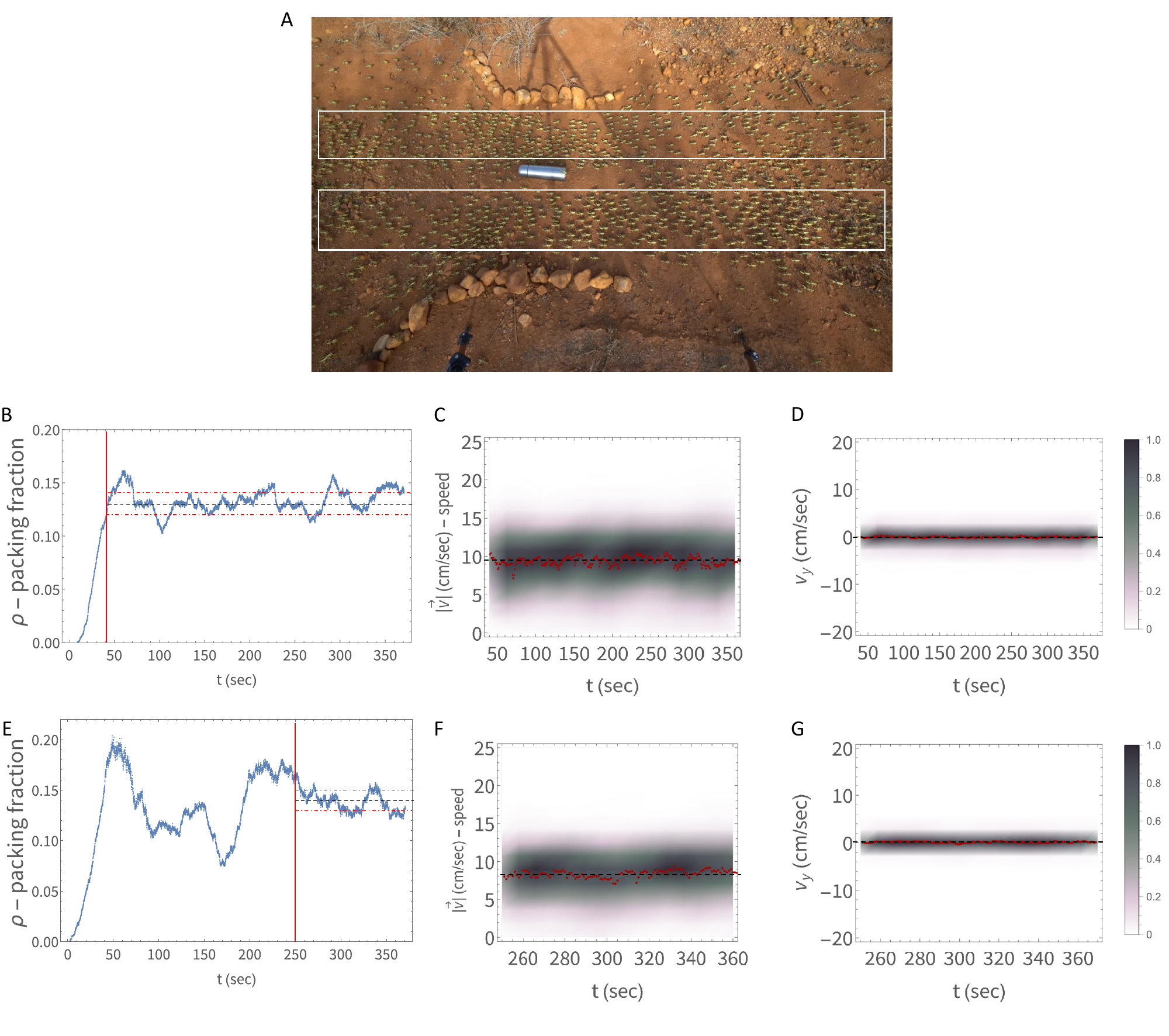}
\caption{\label{Steady}
(A) A frame from video 1. Each frame is 3840x2160 pixels where here the corners of the top strip are at (40,637) and (845,3800). The corners of the bottom strip are at (40,1064) and (1410,3800). The coordinates are measured from the upper left corner of the frame. (B) The density of the locusts as a function of time in the bottom strip normalized as packing fraction. Starting from $t_{min}=40 sec$ the flow is steady since the fluctuations around the mean value (the black line) are small (less than $10\%$). The red dashed lines are standard deviations around the mean value.  (C) The speed in the bottom strip for $t>t_{min}$. The black dashed line marks the average speed over the time and the red points are the long wavelength approximation that shows the fluctuations around the average value. (D) The vertical component of the velocity $v_y$ in the bottom strip. The black dashed line is the average and the red points are the fluctuations around the average like in C. (E) The density of the locusts as a function of time in the top strip normalized as packing fraction. Starting from $t_{min}=250 sec$ the flow is steady since the fluctuations around the mean value (the black line) are small (less than $10\%$).  (F) The speed in the bottom strip for $t>t_{min}$. The black dashed line marks the average speed over the time and the red points are the long wavelength approximation that shows the fluctuations around the average value. (G) The vertical component of the velocity $v_y$ in the top strip. The black dashed line is the average and the red points are the fluctuations around the average like in F.
}
\end{figure}

We start the analysis of the steady flow by considering two long strips on the sides of the obstacle. The vertical boundaries were chosen to minimize edge effects and were thus taken to be one body length from all the edges (see Fig.~\ref{Steady}A for video 1). The strips are taken to be large enough to average out fluctuations in density over time. The horizontal boundaries were chosen to minimize tracking artifacts of appearing and disappearing individuals near the boundaries. Therefore, they do not coincide with the frame boundaries. In addition, the recordings were rotated, so that the horizontal direction is in the direction of the mean velocity.

The packing fraction as a function of time is shown for the bottom strip of video 1 in Fig.~\ref{Steady}B and for the top strip in Fig.~\ref{Steady}E. The packing fraction here is the density normalized by the maximal possible density of locusts globally for the entire strip. An alternative way to define packing fraction by $\alpha$-shapes is given below (Fig.~\ref{Packing Fraction}). We see that starting from a time of $t=40 sec$ in the bottom strip the fluctuations of the density around the average are small and that the packing fraction  $\rho=0.13\pm 0.01$ in the bottom strip (and $\rho=0.14\pm 0.01$ in the top strip). In order to show that we have a steady flow, we must check the velocity as well. For this purpose, we extract the velocity by taking the mean slopes for 25 frames (1 sec) in both the x and y displacements as a function of time. Then we can calculate the speed $v=\sqrt{v_x^2+v_y^2}$ which fluctuates around a constant value (see the dashed black line in Fig.~\ref{Steady}C). After binning the speed of all individual trajectories of the strip into 25 frames (which is consistent with the smoothing of the velocity in the long wavelength approximation), we get steady values of the speed in both strips (the red points in Fig.~\ref{Steady}C), that describe the long wavelength fluctuations around the average values. In the bottom strip, the average speed is: 
\be
v=9.38\pm 0.53 cm/sec,
\ee
(and $v=8.08\pm 0.57 cm/sec$ in the top strip) where we take the average length of the locust

($4\pm 0.3$ cm \footnote{The measurement was done using a caliper on 10 different locusts}) 
\setcounter{footnote}{1}
and consider only $t>t_{min}$.
In the same way one can check that the direction of the velocity is steady as well by considering $v_y$ (see Fig.~\ref{Steady}D and G) on the same scale relative to the speed. We see that, on average, $v_y$ is very small compared to the speed, so that

\be
v_y/v=-8.5\cdot10^{-3}\pm1\cdot 10^{-2}
\ee
in the bottom strip (and $v_y/v=6.2\cdot10^{-3}\pm2\cdot10^{-2}$ in the top strip).

\clearpage

\section{Results for a second strip}
\label{AppendixTopStrip}
\begin{figure}[htb]
\centering
\includegraphics[width=0.7\linewidth]{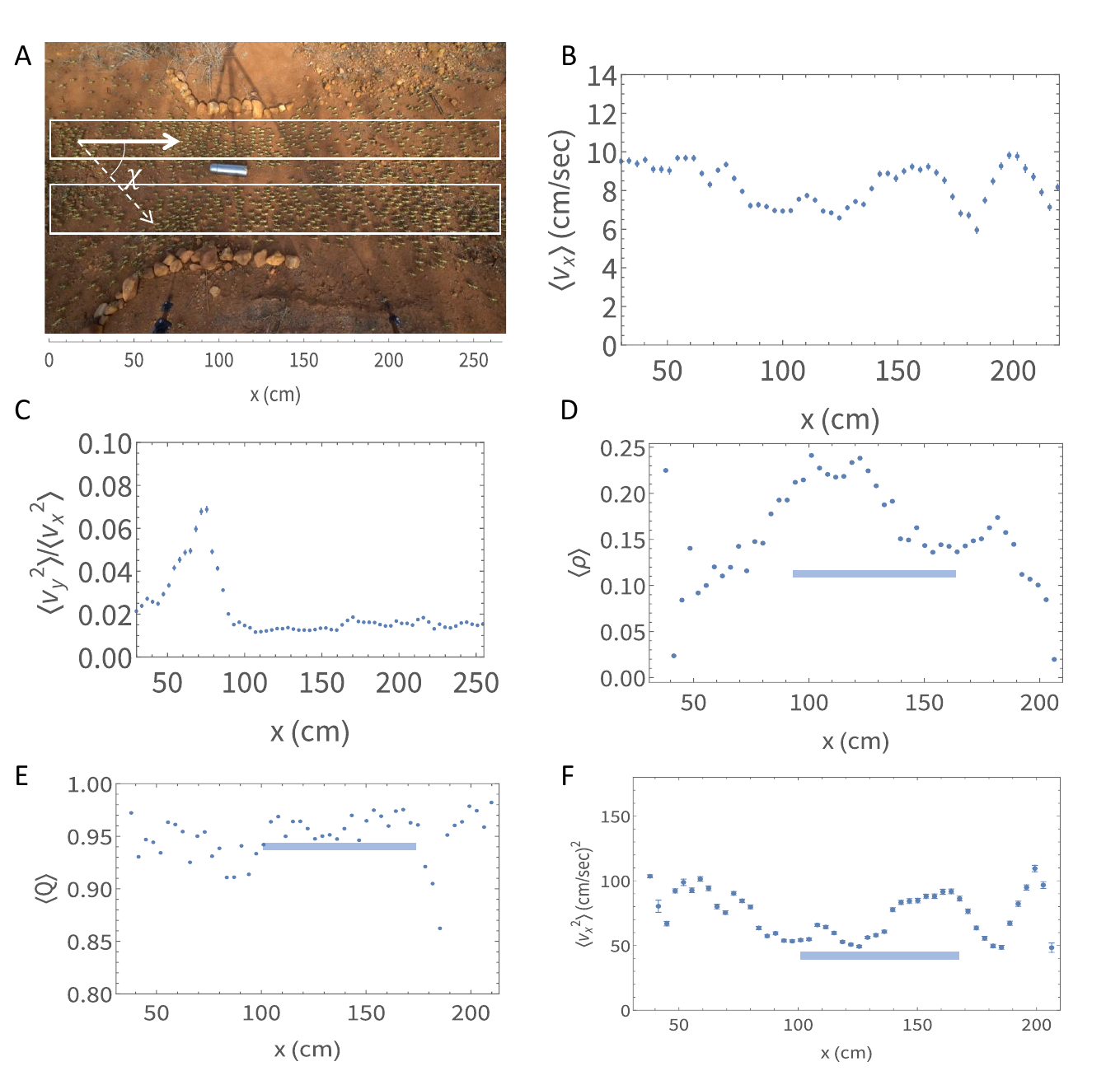}
\caption{\label{SecondB1}
(A) A snapshot of video 1 with the areas took for analysis marked by white boxes. All panels show analysis of quantities averaged over time for the top strip for which the definition of the alignment angle $\chi$ and horizontal distance ruler are marked (in cm). (B) The horizontal component of the velocity $v_x$ as a function of the horizontal coordinate $x$ (average over time).(C) the ratio of the velocity components squared as a function of $x$. (D) The density $\langle \rho \rangle$ (packing fraction) as a function of the horizontal coordinate $x$. (E) The alignment order parameter as a function of $x$. The horizontal blue line denotes the segment where the alignment is maximal $ \langle Q \rangle >0.95$. (F) The horizontal component of the velocity $v_x$ squared as a function of $x$.  }
\end{figure}

\begin{figure}[htb]
\centering
\includegraphics[width=0.5\linewidth]{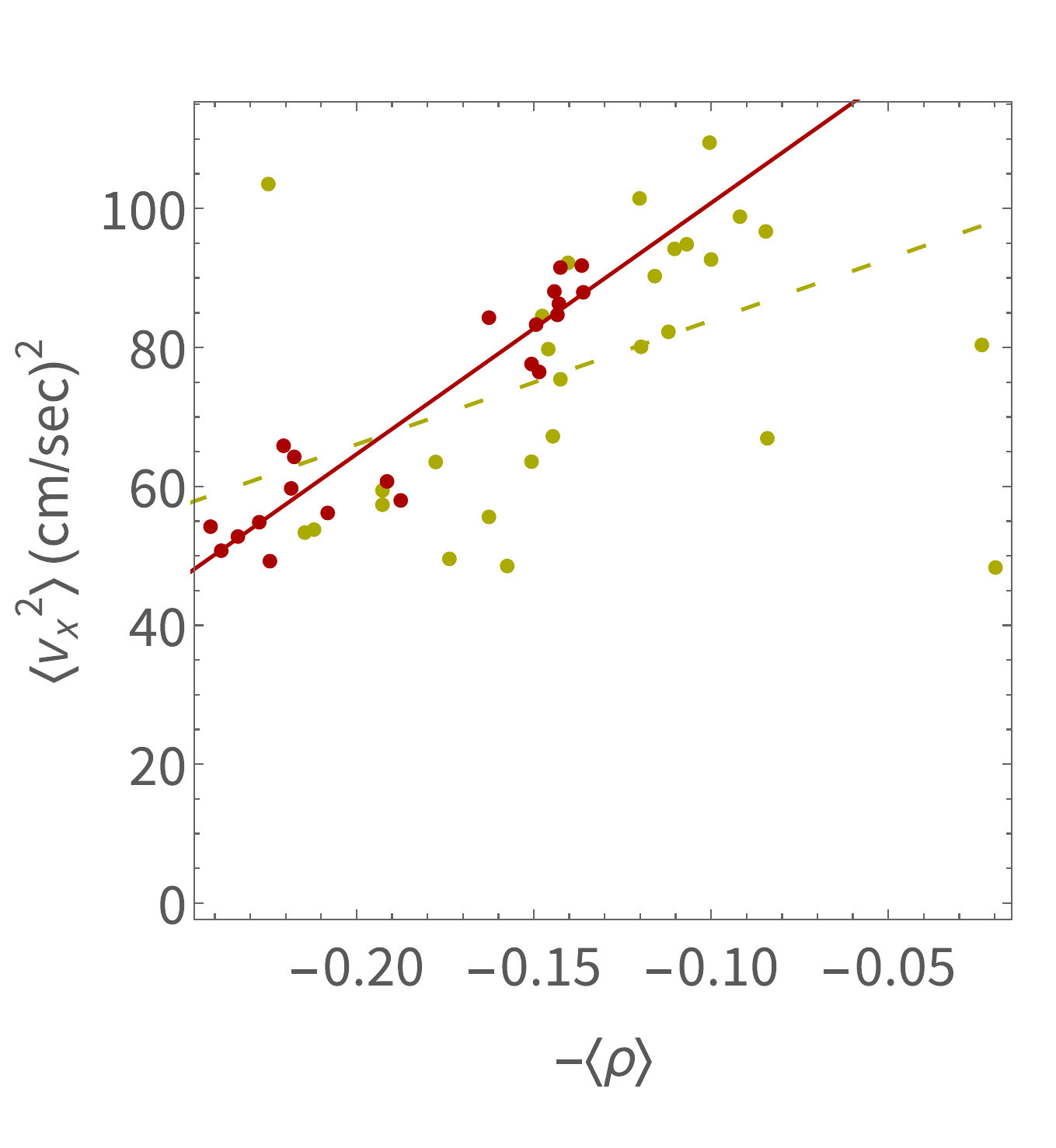}
\caption{\label{SecondB2}  $v_{x}^2$ as a function of -$ \langle \rho \rangle$ for the top strip. The red points correspond to strong alignment $ \langle Q \rangle >0.95$ (from Fig.~\ref{SecondB1}) and the yellow points are the rest. The linear fit of the points with strong alignment is in red and $R^2=0.88$. When we include all the points in the fit we get the dashed yellow line with $R^2=0.28$}
\end{figure}

Here we present in Figs.~\ref{SecondB1} and~\ref{SecondB2} the results of the analysis for the top strip of video 1 that were obtained in a similar way to the bottom strip (see Fig.~\ref{Second}). The results along the horizontal direction along the strip were obtained after averaging over long time segment that was obtained in Fig.~\ref{Steady}E-G.  

\clearpage

\section{Global packing fraction}
\label{AppendixGlobalPacking}
\begin{figure}[htb]
\centering
\includegraphics[width=0.7\linewidth]{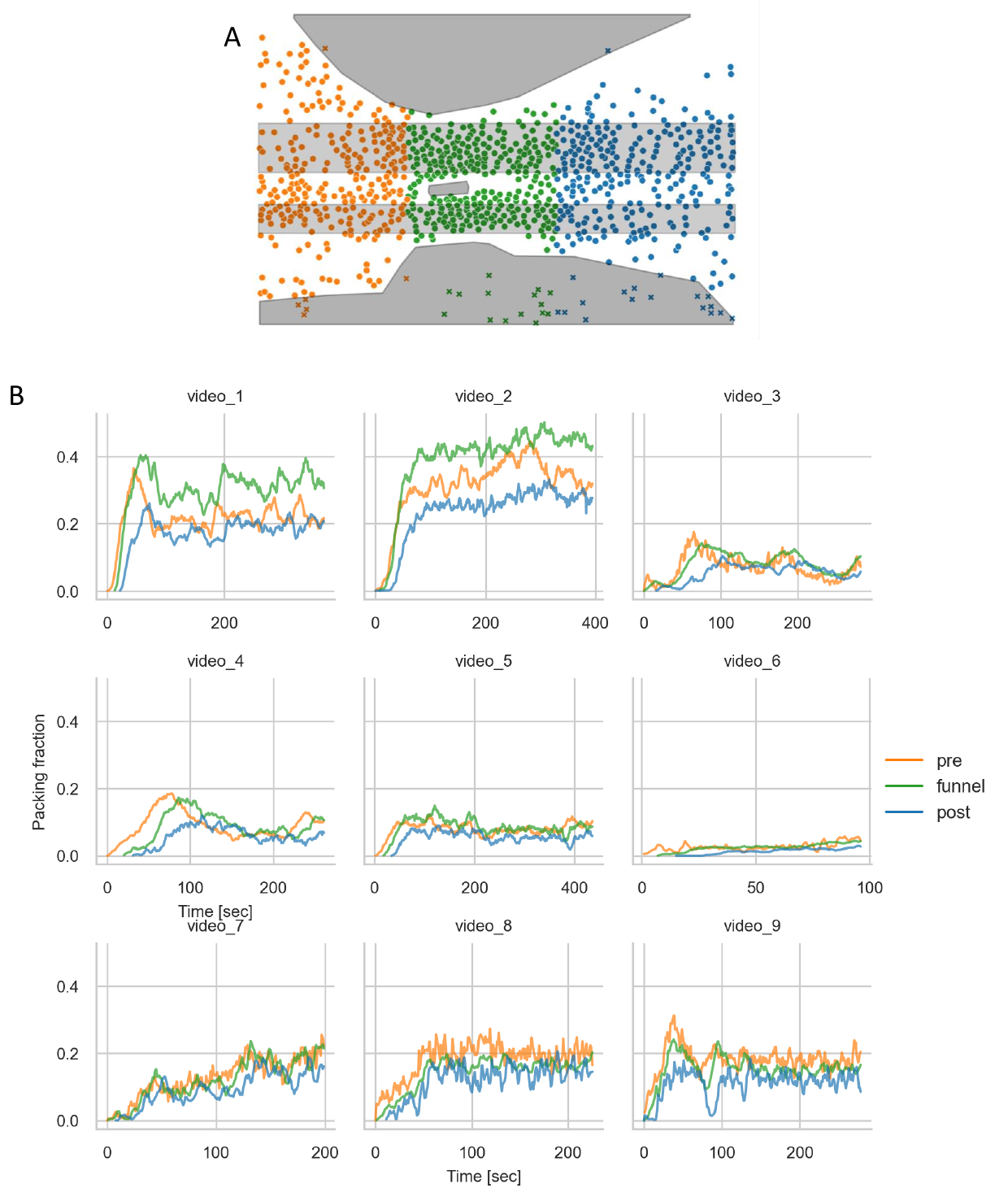}
\caption{(A) For each recording, horizontal strips were assigned at a distance of one body length from obstacles. Three vertical segments (Pre, Funnel, Post) were assigned based on changes in path width, and hence density, along the axis of movement (left to right). Locusts within obstacles were disregarded (crosses).
(B) Global packing fractions for all analyzed recordings over time. \label{Packing Fraction}
 }
\end{figure}

Here we calculate the global packing fraction using $\alpha$-shapes and compare it between three segments in all recordings: Before the funnel - ``Pre'', the funnel itself - ``Funnel'', and after the funnel - ``Post''.  
 For each strip and frame an $\alpha$-shape with a pre-defined $\alpha = 15$ was calculated according to the algorithm that is described in ~\cite{liao2021grid}. We can understand it as constructing a point set boundary by rolling a circle with radius $\alpha$ outside the edge of the set. The alpha parameter value was set to assure robust detection of locust groups without fragmentation (too many subgroups) or streching (including outliers) across recordings via manually inspecting the resulting alpha shapes. Then we calculated the ratio of total area occupied by locusts to the total area of the $\alpha$-shape by fitting an ellipse through each locust’s bounding box. We will refer to this value as the global packing fraction. Global packing fractions were calculated for each of the three segments along the x-axis (shown in figure \ref{Packing Fraction}).
\section{Quadrants}
\label{AppendixQuadrants}
\begin{figure}[h]
\centering
\includegraphics[width=0.4\linewidth]{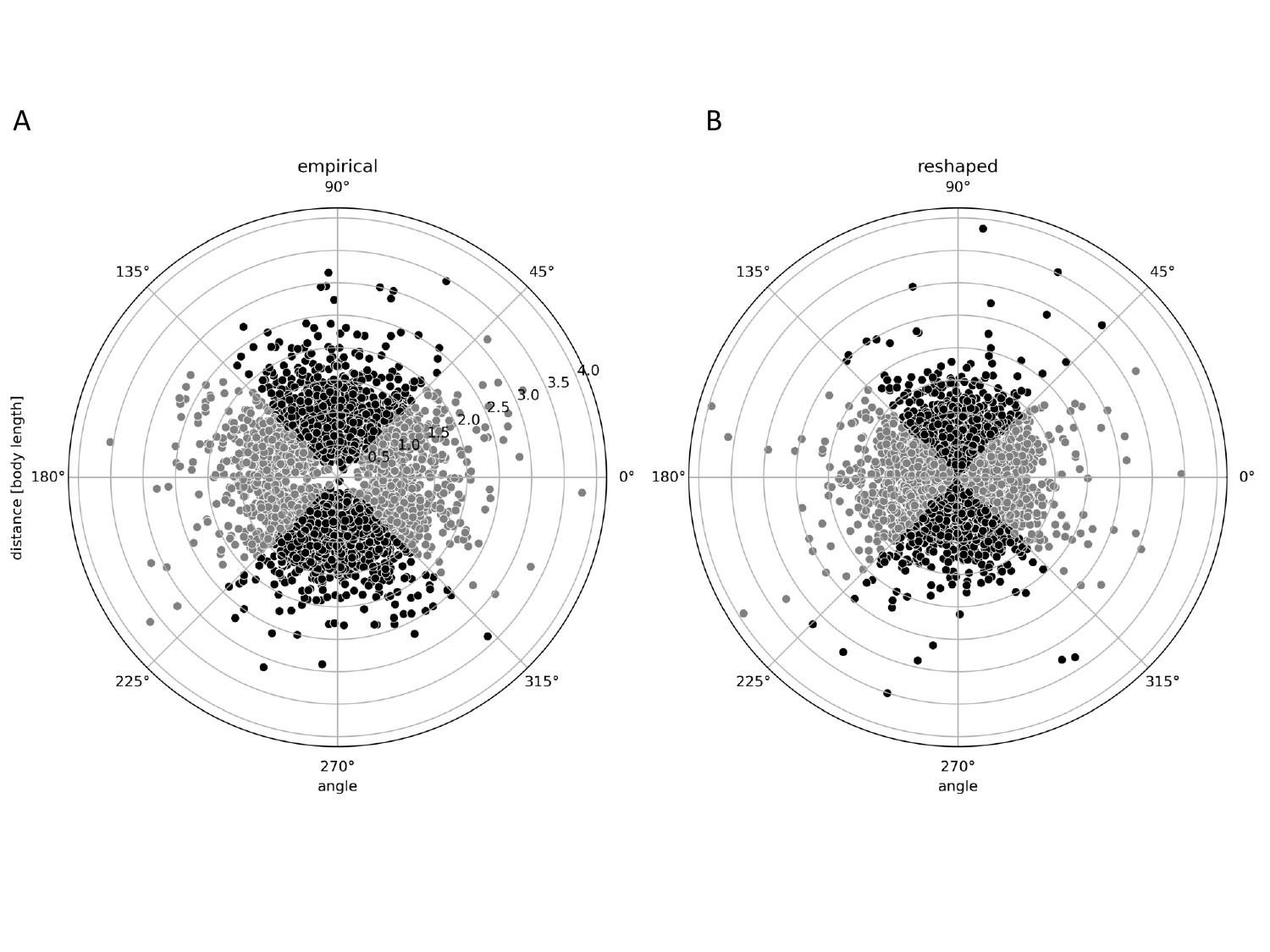}
\caption{The quadrants used to calculate the isotropic distribution. Each dot represents the position of a nearest neighbor relative to the focal individual. The angle bins are defined as 315°-45°, 45°-135°, 135°-225°, and 225°-315°. The grey quadrants represent the back (left) and front (right) of the focal locust, while the black quadrants represent the sides. As a measurement for anisotropy, the number of nearest neighbours on the sides (black) was divided by the total number of neighbours separately for empirical data (A) and rescaled data (B). \label{quadrants}
 }
\end{figure}

Here we present the definition of the angular quadrants that are used to quantify the amount of anisotropy (Fig.~\ref{quadrants}).

\vspace{3 cm}

\end{document}